\documentstyle[epsf]{article}

\makeatletter
\newcommand{\sw}[1]{{\mbox{\scriptsize #1}}}
\newcommand{\bra}[1]{\mbox{$\left\langle{#1}\right|$}}
\newcommand{\ket}[1]{\mbox{$\left|{#1}\right\rangle$}}
\newcommand{\bold}[1]{\mbox{\boldmath$#1$}}

\newcommand{\shamelessUWplugsmall}{%
\font\fortssbx=cmssbx10 scaled \magstep1
\hbox to \hsize{\epsfysize=20pt\epsffile{uwlogo.eps}
\hfil\vbox{\hbox{\vbox to 20pt{\vfill\hbox{\hfil\fortssbx
University of Wisconsin - Madison\hfil}\vfill}}}\hfil}%
\vspace*{0.3in}
}

\newcommand{\shamelessUWpluglarge}{%
\font\fortssbx=cmssbx10 scaled \magstep2
\hbox to \hsize{\epsfysize=30pt\epsffile{uwlogo.eps}
\hfil\vbox{\hbox{\vbox to 30pt{\vfill\hbox{\hfil\fortssbx
University of Wisconsin - Madison\hfil}\vfill}}}\hfil}%
\vspace*{0.5in}
}

\def\printpagenum{\ifnum\thepage=0 \else \thepage \fi}
\def\landscapemode{%
	\renewcommand{\baselinestretch}{1.1}
	\twocolumn\sloppy\flushbottom\parindent 2em
	\leftmargini 2em\leftmarginv .5em\leftmarginvi .5em
	\oddsidemargin 0in	\evensidemargin 0in
	\columnsep .4in	\footheight 0pt
	\textwidth 10in	\topmargin  -.4in
	\headheight 0pt \topskip 0in
	\textheight 6.9in \footskip 30pt
	\hoffset -.5in \voffset -.25in
	\def\@oddfoot{\hfil\printpagenum\hfil\addtocounter{page}{1}
		\hspace{\columnsep}\hfil\printpagenum\hfil}
	\let\@evenfoot\@oddfoot	\def\@oddhead{}	\def\@evenhead{}
	\def\draftstamp{\draftstamptwoup}
	\def\shamelessUWplug{\shamelessUWplugsmall}
	\newcommand{\figuresize}{0.5}
	\def\TitlePageNoNumber{}}

    \def\eqnarray{\stepcounter{equation}\let\@currentlabel=\theequation
    \global\@eqnswtrue
    \global\@eqcnt\z@\tabskip\@centering\let\\=\@eqncr
    $$\halign to \displaywidth\bgroup\@eqnsel\hskip\@centering
      $\displaystyle\tabskip\z@{##}$&\global\@eqcnt\@ne
       \hfil${{}##{}}$\hfil
      &\global\@eqcnt\tw@ $\displaystyle\tabskip\z@{##}$\hfil
       \tabskip\@centering&\llap{##}\tabskip\z@\cr}

\def\section{\@startsection{section}{1}{\z@}{-1.5ex}{1.5ex}%
{\large\bf}}
\def\subsection{\@startsection{subsection}{2}{\z@}{-1.2ex}{1.2ex}%
{\normalsize\bf}}
\def\subsubsection{\@startsection{subsubsection}{3}{\z@}{-1ex}{1ex}%
{\small\bf}}

\def\preprintnum#1{\hspace*{\fill #1}\par}

\def\title#1{\TitlePageNoNumber\vskip2.5pc
    \begin{center}
    {\LARGE #1 \par}%
    \end{center}
    \vskip 1.5em}
\def\author#1#2{\renewcommand{\thefootnote}{\fnsymbol{footnote}}
	\begin{center}
    {\large \lineskip .5em
    #1\footnote{e-mail: {\tt #2}} \par}
    \end{center}
    }
\def\institution#1{\begin{center}
    {\large \lineskip .5em \em #1 \par}
    \end{center}
    \vskip 1.5em}
\def\abstract#1{{\small
    \begin{center}%
    {\bf Abstract\vspace{-.5em}\vspace{\z@}}%
    \end{center}%
    \quotation #1 \endquotation\newpage}}

\newlength{\minuswidth}
\newlength{\digitwidth}
\newlength{\pointwidth}
\def\citen#1{%
\edef\@tempa{\@ignspaftercomma,#1, \@end, }
\edef\@tempa{\expandafter\@ignendcommas\@tempa\@end}%
\if@filesw \immediate \write \@auxout {\string \citation {\@tempa}}\fi
\@tempcntb\m@ne \let\@h@ld\relax \let\@citea\@empty
\@for \@citeb:=\@tempa\do {\@cmpresscites}%
\@h@ld}
\def\@ignspaftercomma#1, {\ifx\@end#1\@empty\else
   #1,\expandafter\@ignspaftercomma\fi}
\def\@ignendcommas,#1,\@end{#1}
\def\@cmpresscites{%
 \expandafter\let \expandafter\@B@citeB \csname b@\@citeb \endcsname
 \ifx\@B@citeB\relax 
    \@h@ld\@citea\@tempcntb\m@ne{\bf ?}%
    \@warning {Citation `\@citeb ' on page \thepage \space undefined}%
 \else
    \@tempcnta\@tempcntb \advance\@tempcnta\@ne
    \setbox\z@\hbox\bgroup 
    \ifnum\z@<0\@B@citeB \relax
       \egroup \@tempcntb\@B@citeB \relax
       \else \egroup \@tempcntb\m@ne \fi
    \ifnum\@tempcnta=\@tempcntb 
       \ifx\@h@ld\relax 
          \edef \@h@ld{\@citea\@B@citeB}%
       \else 
          \edef\@h@ld{\hbox{--}\penalty\@highpenalty \@B@citeB}%
       \fi
    \else   
       \@h@ld \@citea \@B@citeB \let\@h@ld\relax
 \fi\fi%
 \let\@citea\@citepunct
}
\def\@citepunct{,\penalty\@highpenalty\hskip.13em plus.1em minus.1em}%
\def\@citex[#1]#2{\@cite{\citen{#2}}{#1}}%
\def\@cite#1#2{\leavevmode\unskip
  \ifnum\lastpenalty=\z@ \penalty\@highpenalty \fi 
  \ [{\multiply\@highpenalty 3 #1
      \if@tempswa,\penalty\@highpenalty\ #2\fi 
    }]\spacefactor\@m}
\renewcommand{\cite}[1]{[\citen{#1}]}

\makeatletter\input{art12.sty}
	\renewcommand{\baselinestretch}{1.1}
	\def\draftstamp{\draftstampsubmit}
	\def\shamelessUWplug{\shamelessUWpluglarge}
	\newcommand{\figuresize}{0.5}
	\def\TitlePageNoNumber{\thispagestyle{empty}}

\newcommand{\lambar}{{\overline \Lambda}}
\newcommand{\lamtilde}{\tilde{\Lambda}}
\newcommand{\vecr}{\bold{r}}
\newcommand{\vecp}{\bold{p}}
\newcommand{\phinr}{\phi^\sw{NR}}
\newcommand{\phisr}{\phi^\sw{SR}}

\begin{document}

\shamelessUWplug

\preprintnum{MAD/PH/859}
\preprintnum{hep-ph/9504425}

\title{\Large Comparison of potential models through HQET}

\author{James F. Amundson}{amundson@phenom.physics.wisc.edu}

\institution{Department of Physics, University of Wisconsin,\\
Madison WI 53706}

\abstract{
I calculate heavy-light decay constants in a nonrelativistic potential
model.  The resulting estimate of heavy quark symmetry breaking conflicts
with similar estimates from lattice QCD.  I show that a semirelativistic
potential model eliminates the conflict.  Using the results of heavy quark
effective theory allows me to identify and compensate for shortcomings in
the model calculations in addition to isolating the source of the
differences in the two models.  The results lead to a rule as to where the
nonrelativistic quark model gives misleading predictions.  }

\section{Introduction}
The nonrelativistic quark model is one of the oldest and most successful
models of hadronic physics.  This success is somewhat puzzling in that it
persists even when the model is applied to light quark hadrons, where the
dynamics are dominantly relativistic.  Perhaps more puzzling is that
relativistic corrections to the nonrelativistic quark model do not to
substantially improve the model's predictions for spectra
\cite{LucRupSchob}.  Some (but not all) of the ideas of the nonrelativistic
quark model for heavy-light systems gain a stronger theoretical basis
through heavy quark effective theory (HQET).  In this work I show how the
nonrelativistic quark model can be used in conjunction with HQET to
calculate heavy-light decay constants.  By doing the same calculation with
a semirelativistic potential model, I show how relativistic extensions of
the simple quark model can make a dramatic improvement in some types of
calculations.  This, in turn, indicates which nonrelativistic quark model
calculations should not be trusted.

Before turning to the model calculations it is important to understand what
HQET tells us about decay constants, since HQET provides the only results
that follow directly from QCD.  The application of the ideas of HQET to
heavy-light decay constants preceded the development of the effective
theory itself.  The nonrelativistic quark model led to the prediction that
heavy-light decay constants follow the scaling behavior
\cite{AzFranKhoz}
\begin{equation}
	f_M \propto \frac{1}{\sqrt{m_M}}.
	\label{eq:prehistory}
\end{equation}
Later, Shifman and Voloshin \cite{VolShif} and, separately, Politzer and
Wise \cite{PolWise} calculated the leading-logarithmic corrections to
Eq.~(\ref{eq:prehistory})
\begin{equation}
	{f_B\over f_D} = \left[{\alpha_s(m_c)\over\alpha_s(m_b)}\right]^{6/25}
		\sqrt{m_D\over m_B}
	\label{eq:dawnoftime}
\end{equation}
in a model-independent manner, {\em i.e.}, following directly from QCD in
the limit where the heavy quark mass goes to infinity while the the QCD
scale remains fixed.

The above relation is of both theoretical and practical interest.
Theoretically, Eq.~(\ref{eq:dawnoftime}) is interesting because it is a
model-independent prediction of QCD in a well-defined limit.  Practically,
it is interesting because $f_B$ is an input to other calculations, such as
$B^0$-$\bar B^0$ mixing.  Unfortunately, a direct measurement of $f_B$
through leptonic decay will be extremely challenging because of the very
small branching ratio and difficult signature.  A measurement of $f_D$, on
the other hand, is much more feasible.  In fact, measurements of $f_{D_s}$,
which is related to $f_D$ by flavor SU(3), are already available
\cite{fDsWA75,fDsCLEO,fDsBES}, albeit with large errors.

Unfortunately, in the real world the bottom and, particularly, the charm
quark masses are quite finite compared to the QCD scale.  It is therefore
necessary to consider the finite-mass corrections to
Eq.~(\ref{eq:dawnoftime}).  The predictive power of the effective theory
vanishes when the leading-order finite mass corrections the decay constants
are included.  This means the size of the corrections must be estimated
using lattice QCD or some model.  This problem has been studied extensively
on the lattice \cite{Hashimoto,Abada,Bernard,Baxter}, where results
indicate that the corrections to the heavy quark limit for $f_B$ are
$\cal{O}(20\%)$, which corresponds to a subleading heavy quark term of size
$(1 \mbox{~GeV})/m_Q$.  QCD sum rules \cite{Ball} are consistent with these
lattice results.  The large correction is surprising when compared to what
one would naively expect from the nonrelativistic quark model, something
like $(0.3 \mbox{~GeV})/m_Q$.  Naive estimates can miss factors of three,
of course.  It is necessary to do an explicit calculation to see that the
nonrelativistic quark model really conflicts with the lattice calculations.

This work uses two simple potential models to explicitly calculate decay
constants in the heavy quark limit and beyond.  The first, hereafter
referred to as the ``nonrelativistic quark model'' is based on the
Hamiltonian of the Isgur-Scora-Grinstein-Wise (ISGW) model \cite{ISGW} in
the heavy quark limit.  This model is very simple, in contrast to lattice
methods, which are rigorous, but also exceedingly complicated.  Even if the
lattice is able to provide precise answers to the structure of hadrons, it
is useful to find simple pictures which describe the important physics.
The second model, the ``semirelativistic quark model'' is an simple
generalization of the first.  The difference is substitution of the
relativistic form for the kinetic energy for the nonrelativistic form used
in the ISGW model.  This simple change involves subtleties which are
discussed in the body of the text.

One might reasonably ask, given a willingness to use these models for
calculations, why bother with HQET at all? There are several reasons. The
first is that HQET provides some checks on the calculation. At subleading
order, HQET does show that there is a term missing from the model
calculation. Fortunately, it is a term that may be added by hand. A
second reason is that HQET allows the inclusion of radiative corrections in
a rigorous manner. Finally, and perhaps most importantly, HQET provides a
{\em detailed} way to compare models. When models differ in their
predictions, it is desirable to isolate the regions in which they differ.
Unfortunately, when one takes apart two different models to compare, it
becomes a matter of comparing apples and oranges. By calculating
nonperturbative matrix elements that arise in HQET, the two models can be
compared in a physically meaningful way.

The next section reviews the HQET predictions for decay constants to
subleading order in $1/m_Q$.  The following sections describe the
calculations in the nonrelativistic and semirelativistic models.  I then
compare the results of the two models and discuss the implications for
other nonrelativistic quark model calculations.  The appendix describes the
numerical methods I used to do the calculations.

\section{HQET for meson decay constants}
The heavy quark effective Lagrangian \cite{Georgi,EichHill},
\begin{equation} \label{eq:LHQ}
	{\cal L}_\sw{HQ} = \bar h_v D\cdot v h_v,
\end{equation}
is by now well known.  For a review which includes an extensive discussion
of decay constants, see Ref.~\cite{NeubertReview}.  The spin and heavy
quark mass symmetries of the heavy quark limit are manifested by the lack
of gamma matrices and masses in Eq.~(\ref{eq:LHQ}).  The usual definition
of the pseudoscalar decay constant, $f_M$, of a $Q\bar q$ meson $M$ with
four-momentum $p$ is
\begin{equation}
	\bra{0}A_\mu\ket{M(p)} = i f_M p_\mu,
	\label{eq:fMdef}
\end{equation}
where $A_\mu$ is the axial current.  Throughout this work $M~(M^*)$
represents a heavy-light pseudoscalar (vector) meson with a heavy quark $Q$
and a light antiquark $\bar q$.  Using the symmetries of
Eq.~(\ref{eq:LHQ}), one can see that in the heavy quark limit
\begin{equation}
	f_M\sqrt{m_M} = F,
	\label{eq:Fdef}
\end{equation}
where $F$ is a universal dimensionful parameter of QCD.  This parameter
depends on the nonperturbative sector of QCD, so it is not currently
calculable from first principles.  (It is calculable on the lattice in
principle.) In the symmetry limit, {\em i.e.,} when the bottom and charm
quarks are taken to be infinitely massive, the decay constants of the $D$,
$D^*$, $B$ and $B^*$ are determined by $F$.

The discussion so far ignores radiative corrections.  When the
leading-logarithmic radiative corrections to the axial current in the heavy
quark limit are included, the result becomes
\begin{equation}
	f_M\sqrt{m_M} = \left[{\alpha_s(\mu)\over \alpha_s(m_Q)}\right]
	^{\frac{2}{\beta_0}}F(\mu),
	\label{eq:Fmudef}
\end{equation}
where $\beta_0 = (33-2n_f)/3$. The general form of the result is
$C(\mu)F(\mu)$, where $C(\mu)$ is the perturbative coefficient to the
low-energy parameter $F(\mu)$. Since physics does not depend on the choice
of scale, the $\mu$-dependence of the product must vanish.

At subleading order in $1/m_Q$ and including leading-log radiative
corrections, the Lagrangian grows \cite{FalGrin,FalGrinLuk,EichHill2}:
\begin{equation}
	{\cal L}_\sw{QCD} = {\cal L}_\sw{HQ} + {1\over2m_Q}
	\bar h_v D^2 h_v + {1\over4m_Q}
	\left[\alpha_s(\mu)\over\alpha_s(m_Q)\right]^{-{3\over\beta_0}}
	\bar h_v g_s\sigma^{\mu\nu}G_{\mu\nu} h_v +
	{\cal O}\left(\Lambda_\sw{QCD}^2\over m_Q^2\right).
\end{equation}
A third term at order $1/m_Q$ whose matrix elements vanish due to the
equations of motion has been omitted.  The first correction term is the
leading part of the kinetic energy of the heavy quark.  Its perturbative
coefficient is unity because of reparameterization
invariance \cite{LukeMano}.  The second correction term arises from the
heavy quark's non-zero chromomagnetic moment.  These terms give rise to
corrections to the decay constant through modifications of the meson wave
function and of the heavy-light current.  When these effects are included,
the simple result in Eq.~(\ref{eq:Fdef}) becomes (ignoring radiative
corrections for simplicity) \cite{NeubertDC}
\begin{equation}
	f_M\sqrt{m_M} = F\left\{1 + \frac{1}{m_Q}\left[G_1 + 2d_MG_2\right] -
	d_M\frac{\lamtilde}{6m_Q}\right\},
	\label{eq:fHQsub}
\end{equation}
where $d_M = +3~(-1)$ for pseudoscalar (vector) mesons.  Here the effect of
the modification to the meson wave function due to the kinetic energy term
and the chromomagnetic term are parameterized by $G_1$ and $G_2$,
respectively.  The finite difference between the heavy {\em quark} momentum
and the heavy {\em meson} momentum gives rise to the final term.  $G_1$,
$G_2$ and $\lamtilde$ are dimensionful parameters of QCD which, like $F$,
cannot be calculated in perturbation theory.

Unlike $G_1$ and $G_2$, the parameter $\lamtilde$ is directly related to
other heavy quark processes. The difference between the mass $m_M$ of a
heavy-light meson and the corresponding heavy quark mass $m_Q$ is
conventionally defined as
\begin{equation}
	\lambar = m_M - m_Q.
	\label{eq:deflambar}
\end{equation}
$\lamtilde$ is related to $\lambar$ by \cite{NeubertReview}
\begin{equation}
	\lamtilde = \lambar - m_q,
	\label{eq:deflamtilde}
\end{equation}
where $m_q$ is the {\em light} quark mass.  Since the current masses of the
up and down quarks (5--10 MeV) are considerably smaller than estimates of
$\lambar$ (typically 300--700 MeV), $\lamtilde$ is usually taken to be
equal to $\lambar$.  A subtlety which arises in the model calculations in
the following sections makes this distinction important.

Including the leading-log radiative corrections to Eq.~(\ref{eq:fHQsub}),
\begin{eqnarray}
	f_M\sqrt{m_M}& = &\left[{\alpha_s(\mu)\over \alpha_s(m_Q)}\right]
	^{\frac{2}{\beta_0}}F(\mu)\left\{1 + \frac{1}{m_Q}\left[G_1(\mu) + 2d_M
	\left[{\alpha_s(\mu)\over \alpha_s(m_Q)}\right]
	^{\frac{3}{\beta_0}}
	G_2(\mu)\right] - \right. \nonumber\\
	&&\left.\frac{\lamtilde}{6m_Q}\left[\frac{16}{\beta_0}\ln
	\frac{\alpha_s(\mu)}{\alpha_s(m_Q)} + d_M\frac{16}{9}
	\left[{\alpha_s(\mu)\over \alpha_s(m_Q)}\right]
	^{\frac{3}{\beta_0}}\right]
	\right\}.
	\label{eq:fHQsubll}
\end{eqnarray}
An important point about the above complicated expression is that the
$G_1(\mu)$ term has the same perturbative coefficient as $F(\mu)$ because
of reparameterization invariance, while the $G_2(\mu)$ term gets a
non-trivial perturbative coefficient.

At this level the results of HQET have lost their beautiful simplicity.
Unfortunately, they have also lost their predictive power because the four
decay constants $f_{D^{(*)}}$, $f_{B^{(*)}}$ are given in terms of four
unknown parameters $F$, $G_1$, $G_2$ and $\lamtilde$.  Of these, only
$\lamtilde$ is obtainable from other heavy-meson processes in principle.
The task for the model calculations is to estimate these parameters.

\section{Nonrelativistic Model Calculation}

In the nonrelativistic quark model, meson decay constants are given by
\cite{WaveFnOrigin1,WaveFnOrigin2,WaveFnOrigin3,WaveFnOrigin4,WaveFnOrigin5}
\begin{equation}
	f_M\sqrt{m_M}=\sqrt{12}|\psi_M(\bold{r}=0)|.
	\label{eq:fQM}
\end{equation}

The ``non-relativistic quark model'' for this paper is a heavy
constituent quark $Q$ bound to a light antiquark $\bar q$ obeying the
Hamiltonian
\begin{equation}
	{\cal H} = {p^2\over2m_q} + {p^2\over2m_Q} - {4\alpha_s\over3r} + ar +
		{8\pi\alpha_s\bold{S_Q}\cdot\bold{S_q}\over3m_qm_Q}\delta^3(\bold{r}).
\end{equation}
The last three terms in the Hamiltonian are the quark-antiquark potential.
The first and third represent the coulomb-like and hyperfine effects of
single gluon exchange, respectively.  The linear term is a phenomenological
spin-independent confining potential.  A more general Hamiltonian would
also include spin-orbit coupling terms.  I have omitted such terms because
all of the calculations in this work involve only $S$-wave states for which
spin-orbit contributions vanish.  Without the hyperfine term, the
Hamiltonian is that of the ISGW model
\cite{ISGW}. With the hyperfine term, the model is closely related to the
updated model of Isgur and Scora (ISGW2) \cite{ISGW2}. It should be noted
that this model differs from ISGW and ISGW2 in that I use exact
(numerical) wave functions, while the others use simple variational wave
functions. While variational wave functions are useful for the
calculations in ISGW and ISGW2, which involve overlaps of wave functions,
they are not appropriate for decay constants, which are sensitive
to the wave function at a single point.

As the heavy quark mass is taken to infinity, expectation values of $p$ and
$\delta^3(\bold{r})$ remain of order of the QCD scale. In this limit the
above Hamiltonian reduces to \cite{me1}
\begin{equation}
	{\cal H_\infty} = {p^2\over2m_q} - {4\alpha_s\over3r} + ar.
	\label{eq:haminf1}
\end{equation}

\begin{figure}
	\hfill\epsfxsize=0.5\hsize\epsffile{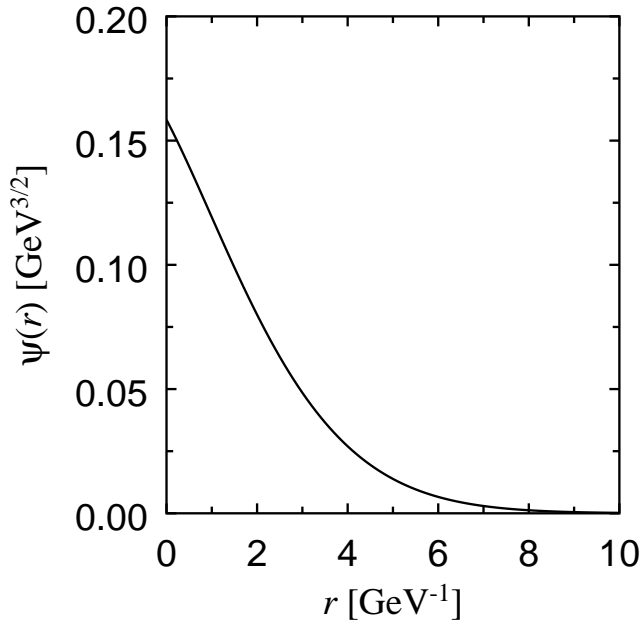}\hfill\hfill

	\caption{Nonrelativistic wave function.}
	\protect\label{fig:nonrel}
\end{figure}

\begin{figure}
	\hfill\epsfxsize=0.5\hsize\epsffile{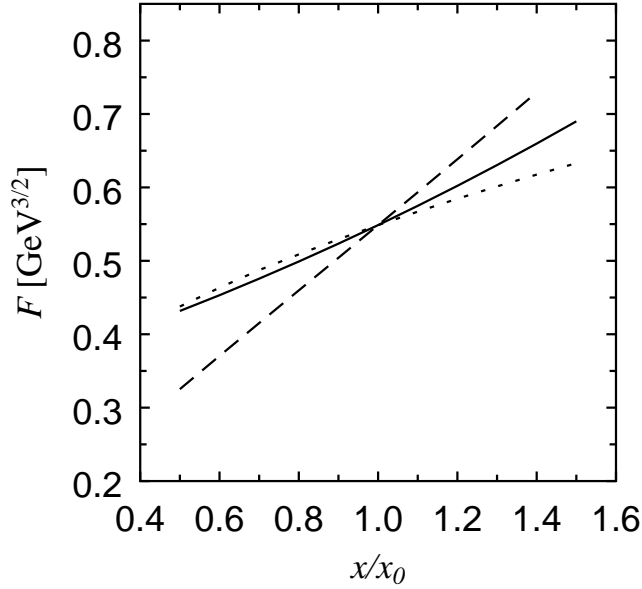}\hfill\hfill

\caption{Parameter dependence of $F$ calculated using the nonrelativistic
model.  The nominal values ($x_0$) are (solid line) $\alpha_s$ = 0.5,
(dashed line) $m_q$ = $0.33$~GeV, and (dotted line) $a = 0.18$~GeV$^2$.}
\protect\label{fig:nonreldep}
\end{figure}

Solving the Schr\"odinger equation
\begin{equation}
	{\cal H_\infty}\phinr_\infty = E \phinr_\infty
	\label{eq:Sch}
\end{equation}
for the ground state wave function and comparing with Eq.~(\ref{eq:fQM})
gives
\begin{equation}
	F = \sqrt{12}|\phinr_\infty(\bold{r}=0)|.
	\label{eq:fnonrel}
\end{equation}
This model calculation explicitly obeys the mass and spin symmetries of the
heavy quark effective theory.  Solving Eq.~(\ref{eq:Sch}) numerically gives
$F=0.55\mbox{~GeV}^{3/2}$.  I have used the parameter values $m_q =
330$~MeV, $\alpha_s = 0.5$, $a = 0.18$~GeV$^2$ from Ref.~\cite{ISGW}.
Figure~\ref{fig:nonrel} displays the calculated wave function.
Unfortunately, numerical calculations such as this tend to obscure the
dependence of the results on the input parameters.  This is particularly
important when trying to establish agreement or disagreement between
different types of calculations.  Figure~\ref{fig:nonreldep} makes the
parameter dependence of the result more explicit by displaying the
dependence of $F$ on the input parameters within $\pm50\%$ of each nominal
value.

In order to calculate the decay constant to subleading order in $1/m_Q$,
it is necessary to reintroduce the heavy quark kinetic energy and hyperfine
interactions to the Hamiltonian. The wave function can be written in an
expansion in powers of $1/m_Q$ as follows
\begin{equation}
	\phinr_M = \phinr_\infty + {1\over m_Q} (\phinr_\sw{KE}+
		d_M\phinr_\sw{HF}) + {\cal
		O}\left({1\over m_Q^2}\right).
\end{equation}
The functions $\phinr_\sw{KE}$ and $\phinr_\sw{HF}$ arise from the effects
of the kinetic energy and spin-spin hyperfine terms, respectively.  They
have been defined to be independent of $m_Q$.  Note that
$4\langle\bold{S}_c\cdot\bold{S}_d\rangle = (-3,1)$ for $(M,M^*)$ is
$-d_M$, which was defined in the previous section.  Simply solving the
Schr\"odinger equation including the $1/m_Q$ terms leads to contributions
of the subleading-mass terms to all orders in $1/m_Q$.  The functions
$\phinr_\sw{KE}$ and $\phinr_\sw{HF}$ can be isolated using perturbation
theory, where
\begin{equation}
	\phinr_\sw{KE}(\vecr) = \sum_{n\neq\infty} {\phinr_n(\vecr)\over E_n -
		E_\infty}\int
		d^3\vecr'\, \phi_n^*(\vecr') \frac{\nabla^2}{2} \phinr_\infty(\vecr')
	\label{eq:kepert}
\end{equation}
and
\begin{equation}
	\phinr_\sw{HF}(\vecr)=
		\frac{1}{4}
		\sum_{n\neq\infty} {\phinr_n(\vecr)\over E_n - E_\infty}\int
		d^3\vecr'\, \phi_n^*(\vecr') \frac{8\pi\alpha_s}{3m_q}\delta^3
		(\bold{r}) \phinr_\infty(\vecr').
		\label{eq:sspert}
\end{equation}
In Eqs.~(\ref{eq:kepert}) and (\ref{eq:sspert}) the set of functions
$\phi_\infty$, $\phi_1$, $\phi_2$\ldots represents the complete set of
solutions to Eq.~(\ref{eq:Sch}) with the unperturbed Hamiltonian in
Eq.~(\ref{eq:haminf1}).  In practice it proves easier to numerically find
the piece of the solution to the full Hamiltonian linear in $1/m_Q$ than to
directly apply Eqs.~(\ref{eq:kepert}) and (\ref{eq:sspert}).

The constants $G_1$ and $G_2$ defined in Eq.~(\ref{eq:fHQsub}) are related
to the wave function corrections by
\begin{equation}
	G_1 = {\phinr_\sw{KE}(0)\over\phinr_\infty(0)}
\end{equation}
and
\begin{equation}
	G_2 = {\phinr_\sw{HF}(0)\over2\phinr_\infty(0)}.
\end{equation}
The quark model calculation to order $1/m_Q$ reproduces the form of the
heavy quark result in Eq.~(\ref{eq:fHQsub}) with the exception of the term
proportional to $\lamtilde$, which is absent in the model calculation.
This missing term manifests one of the limitations of the constituent quark
model.  It can be understood as follows: the factor $\lamtilde = \lambar -
m_q$ arises in a process with $q^2 = m_M^2$, which is large compared to
$\Lambda_\sw{QCD}$.  The relevant light quark mass $m_q$ should therefore
be the current quark mass, which means
\begin{equation}
\lamtilde = \lambar - m_q^\sw{current} \simeq \lambar.
\end{equation}
The quark model only knows about constituent quarks, however, which would
give
\begin{equation}
\lamtilde^\sw{quark model} = \lambar - m_q^\sw{constituent} = 0.
\end{equation}
This facet of the quark model calculation is wrong.  Fortunately, the
deficiency can be compensated for by manually including the $\lamtilde$
term.

A more serious problem arises in the calculation of the hyperfine
correction to the wave function ($\phinr_\sw{HF}$). A straightforward
evaluation of the sum in Eq.~(\ref{eq:sspert}) shows that $\phinr_\sw{HF}$
(and consequently $G_2$) diverges. The delta-function potential is too
singular for the Schr\"odinger equation so the wave function at the origin
diverges, even at leading order in perturbation theory. Although it is
possible to regulate this singularity through a variety of methods, the
resulting calculation depends critically on the method chosen. Since the
effect of the perturbation on the wave function (eigenfunction) is infinite,
one might naively expect that the effect of the perturbation on the mass
(eigenvalue) would also be infinite. Then the correction to the heavy
particle mass, which is measurable through the $B$-$B^*$ mass splitting,
could be calculated in the regularized theory and subsequently be used to
fix the regularization parameter by fitting to the measured mass splitting.
Unfortunately, this procedure fails because the effect of the hyperfine
perturbation gives a finite eigenvalue correction, even though it gives an
infinite eigenfunction correction.

I will assume that the hyperfine contribution, and subsequently $G_2$, is
negligible compared to the kinetic energy contribution with the following
justifications: Qualitatively, one can compare the terms in the
Eq.~(\ref{eq:sspert}) sum with the terms in the Eq.~(\ref{eq:kepert}) sum.
In both cases the first terms in the series, {\em i.e.,} the contributions
of the lowest-lying excited states, are larger in magnitude than any other
terms.  Comparing only these first few terms, the kinetic energy
perturbation is much larger than the hyperfine perturbation.  However, the
large-$n$ terms in the hyperfine sum fall only like $1/n$, so the sum
diverges, whereas the terms in the kinetic sum fall quickly enough for the
sum to converge.  From this it seems plausible that an appropriately
regularized calculation will yield $|G_2| < |G_1|$.  Furthermore, two QCD
sum rule calculations \cite{NeubertDC,Ball} give $|G_2| \ll |G_1|$.
Therefore, it is reasonable to assume that the hyperfine interaction can be
neglected in this calculation.

Fortunately, the dominant $G_1$ term is easily calculable in the
nonrelativistic model.  The effect of (re-)introducing the heavy quark
kinetic energy can be incorporated in the usual way by substituting the
reduced mass $m_\sw{red}$,
\begin{equation}
	\frac{1}{m_\sw{red}} = \frac{1}{m_q} + \frac{1}{m_Q},
\end{equation}
for the light quark mass $m_q$.  However, simply making the substitution
introduces corrections to all orders in $1/m_Q$.  This is a problem because
the heavy quark kinetic energy $p^2/(2m_Q)$ is only correct to leading
order in $1/m_Q$.\footnote{Which, of course, leads one to wonder about
higher-order corrections to the light quark kinetic energy, which do not
converge, since typical values of $p$ are of the same order as the
constituent light quark mass $m_q$.  Concerns such as these inevitably lead
to a model with relativistic light-quark kinematics such as the one in the
following section.} Taylor-expanding $\phi(0)$ as a function of
$m_\sw{red}$
\begin{equation}
	\phi(r) = \phi_\infty(r) + (m_\sw{red} -m_q) {\partial \phi_\infty(r)
		\over \partial m_q} + {\cal O}[(m_\sw{red} -m_q)^2],
\end{equation}
yields the following expression for $G_1$
\begin{equation}
	G_1 = {m_q^2 \over \phi_\infty(0)}{\partial \phi_\infty(r)
		\over \partial m_q}.
	\label{eq:nrG1rel}
\end{equation}
Numerically, it is easier to treat the heavy quark kinetic energy as
a perturbation, as described above.

\begin{figure}
	\hfill\epsfxsize=0.5\hsize\epsffile{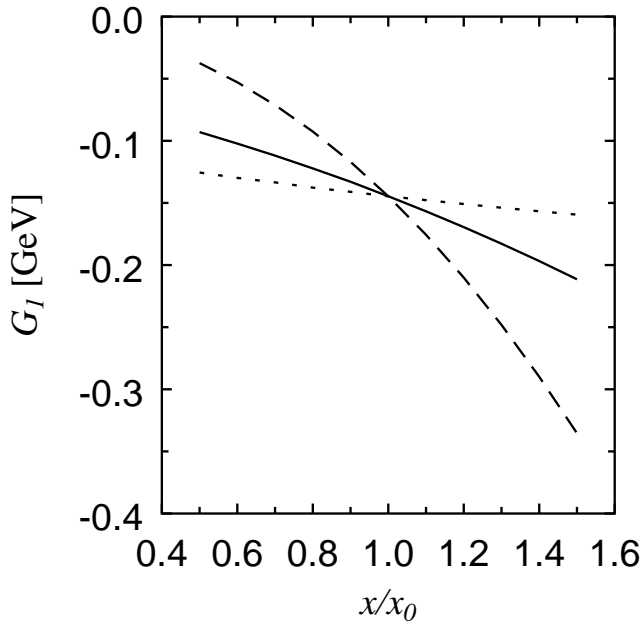}\hfill\hfill

	\caption{Parameter dependence of $G_1$ calculated using the
	nonrelativistic model.  The parameters ($x$) are $\alpha_s$ (solid
	line), $m_q$ (dashed line) and $a$ (dotted line).}

\protect\label{fig:G1nr}
\end{figure}

The numerical calculation gives $G_1~=~-0.14\mbox{~GeV}$.
Fig.~\ref{fig:G1nr} displays the parameter dependence of the calculation,
showing that varying the parameters does not allow for values of $G_1$ much
larger (in absolute value) than 0.3~GeV.

\section{Semirelativistic Model Calculation}
The idea of the semirelativistic model is to remove the most obviously
incorrect part of the nonrelativistic quark model, the nonrelativistic form
of the kinetic energy for the light quark.  Take the nonrelativistic quark
model Hamiltonian in the heavy-quark limit, Eq.~(\ref{eq:haminf1}), and
make the substitution
\begin{equation}
	{p^2\over2m_q} \to \sqrt{p^2+m_q^2}.
	\label{eq:shift}
\end{equation}
The resulting wave equation,
\begin{equation}
	\sqrt{p^2+m_q^2}\:\psi = (E-V)\,\psi,
	\label{eq:spinlesssalp}
\end{equation}
is known as the spinless Salpeter equation \cite{Salpeter}.  It follows
from the full Bethe-Salpeter equation in the spin-independent and
instantaneous-interaction approximation.  The spin-independence is
justified by the heavy-quark limit.  The instantaneous-interaction
approximation is a limitation of the model.  Duncan, Eichten and Thacker
have shown \cite{DunEicTha} that the spinless Salpeter equation produces
wave functions which are very similar to those obtained from lattice
calculations.

If the substitution of the relativistic kinetic energy is the only change
made to the model of the previous section, the potential in
Eq.~(\ref{eq:spinlesssalp}) is
\begin{equation}
	V = -{4\alpha_s\over 3r} + ar.
	\label{eq:pot1}
\end{equation}
Unfortunately, the resulting solution to Eq.~(\ref{eq:spinlesssalp})
diverges at the spatial origin, which results in an infinite value for
$F$ when calculated with Eq.~(\ref{eq:fnonrel}).  One might be tempted to
ascribe this divergence to the phenomenological part of the potential.
However, the divergence depends only on the coulombic part of the potential;
it is independent of the phenomenological linear term.  Wave function
divergence at the spatial origin is actually a general problem affecting
relativistic wave equations.  For example, the solution to the Dirac
equation for the Coulomb potential behaves like
\begin{equation}
	\Psi \sim (2m\alpha r)^{\sqrt{1-\alpha^2}-1}
	\label{eq:diracdiv}
\end{equation}
for small $r$. While the divergence in Eq.~(\ref{eq:diracdiv}) is very weak,
the divergence of the solution to the spinless Salpeter equation with the
potential in Eq.~(\ref{eq:pot1}) is much stronger,
\begin{equation}
	\Psi \sim r^{-{4\alpha_s\over3\pi}},
	\label{eq:ssecouldiv}
\end{equation}
as can be seen with the methods of Ref.~\cite{Durand}.

The singularity in the wave function is clearly related to the singularity
of the $1/r$ potential.  If we consider instead the one-loop single gluon
exchange potential \cite{oneloop1,oneloop2}, the net effect is to replace
the constant value of $\alpha_s$ in Eq.~(\ref{eq:ssecouldiv}) with the
one-loop running value of $\alpha_s(1/\Lambda r)$.  Leaving the
phenomenological linear term unchanged, the potential is now
\begin{equation}
	V = -{4\alpha_s(1/\Lambda r)\over 3r} + ar,
	\label{eq:pot2}
\end{equation}
where
\begin{equation}
	\alpha_s(1/\Lambda r) = { 4\pi \over \beta_0 \ln(1/\Lambda^2 r^2)},
\end{equation}
which has a much milder singularity at the origin.  The resulting wave
function still diverges, but only logarithmically.  Again following a
derivation similar to that in Ref.~\cite{Durand}, one can show that the
small-$r$ behavior of the solution to the spinless Salpeter equation with
the potential in Eq.~(\ref{eq:pot2}) is
\begin{equation}
	\phisr_\infty(r\to0) \sim \left[-\ln(\Lambda r)\right]^{4/3\beta_0}.
	\label{eq:logdiv}
\end{equation}

The physical decay constant is a product of a perturbative coefficient
which depends on a scale $\mu$ with the low-energy parameter $F(\mu)$, as
is shown in Eq.~(\ref{eq:Fmudef}), which is correct to leading-log order.
The logarithmic behavior of the wave function in Eq.~(\ref{eq:logdiv}) is
of the right form to cancel the $\ln(\mu)$ dependence of the perturbative
coefficient that would be obtained if we had only considered single gluon
exchange ({\em i.e.,} the vertex correction) in the perturbative
coefficient in Eq.~(\ref{eq:Fmudef}).  This is as it should be, since the
solution to the wave equation can be considered an infinite series of
single-gluon exchanges.  The full one-loop perturbative calculation also
includes the propagator corrections for the light and heavy quarks, but
these effects are not present in this model. A better model would produce
the full $\ln(\mu)$ dependence of $F(\mu)$.

In the present case, the correct quantity to compare with the
nonrelativistic wave function at the origin is the semirelativistic wave
function at the origin without the logarithmic divergence, which should
cancel with the $\mu$ dependence of the perturbative correction in the
full calculation. Then the quantity
\begin{equation}
	F_0 = \lim_{r\to0}\frac{\phisr_\infty(r)}{\ln(1/\Lambda r)^{4/3\beta_0}}
	\label{eq:F0def}
\end{equation}
should be compared to $F$ as calculated in the nonrelativistic model.

A subtlety arises in the one-loop potential because $\alpha_s(1/\Lambda r)$
diverges for $r\sim\Lambda^{-1}$.  This unphysical behavior arises from the
nonperturbative nature of QCD at long distances, and as such should be
swept into the phenomenological part of the potential.  I have followed the
procedure used by Peskin and Strassler \cite{PesStra} to smoothly turn off
the running of $\alpha_s$ at long distances.  The prescription is to make
the substitution $\Lambda r\to \kappa\mbox{tanh}(\Lambda r/\kappa)$ in the
running of $\alpha_s$.  The results are insensitive to the precise value
of the parameter $\kappa$, which I set to $0.5$ for the results I present
here.

\begin{figure}
	\hfill\epsfxsize=0.5\hsize\epsffile{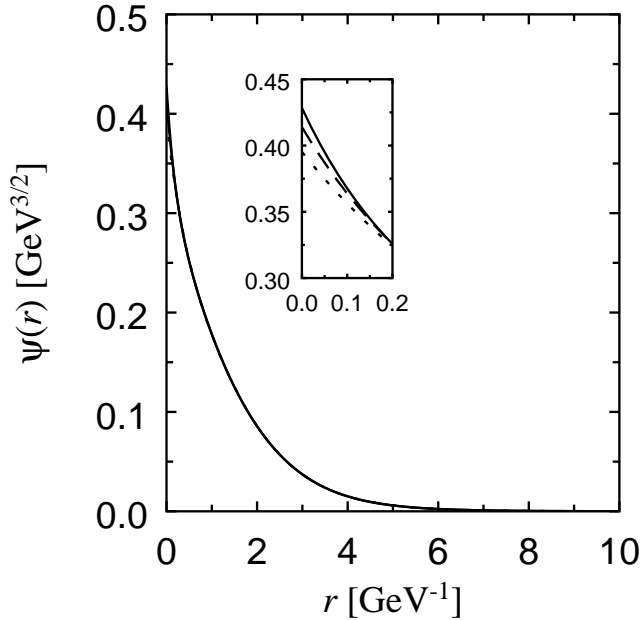}\hfill\hfill

	\caption{Semirelativistic wave function calculated using 10 (dotted
	line), 15 (dashed line) and 20 (solid line)	pseudohydrogenic basis
	functions. The inset shows that the numerical calculation fails to
	converge at the origin, where the wave function diverges
	logarithmically.}
	\protect\label{fig:semirel}
\end{figure}

\begin{figure}
	\hfill\epsfxsize=0.5\hsize\epsffile{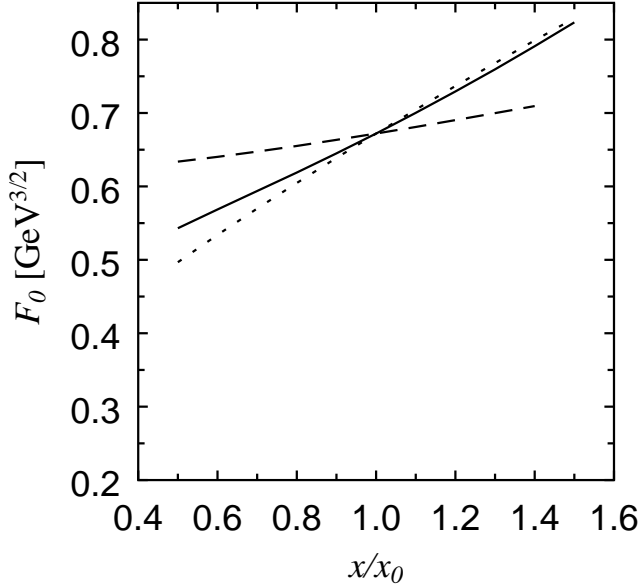}\hfill\hfill

\caption{Parameter dependence of $F_0$ calculated using the
semirelativistic model.  The nominal values ($x_0$) are (solid line)
$\Lambda$ = 0.237, (dashed line) $m_q$ = $0.33$~GeV, and (dotted line) $a$
= 0.18~GeV$^2$.} \protect\label{fig:semireldep}
\end{figure}

Fig.~\ref{fig:semirel} shows the numerical solution to the spinless
Salpeter equation with the one-loop potential.  The values of $m_q$ and $a$
are the same as in the previous section.  There is a subtlety in choosing
$m_q$ in this model.  In one picture the constituent quark mass arises from
the relativistic ``jiggle'' of the light quark in the hadron.  In another
picture, the constituent quark mass arises from chiral symmetry breaking.
Although these schemes are not necessarily mutually exclusive, the former
requires using the current light quark mass in this model, while the latter
requires using the constituent mass.  Here I have chosen the latter
option.  It should be noted, however, that the results do not depend very
strongly on the light quark mass, so choosing the former option would not
qualitatively change the results.  I have chosen $\Lambda=237$~MeV so that
the one-loop potential (Eq.~(\ref{eq:pot2})) is the same as the original
potential (Eq.~(\ref{eq:pot1})) at $r = 1\mbox{ GeV}^{-1}$.
Fig.~\ref{fig:semireldep} displays the sensitivity of resulting value of
$F_0$ to the model parameters. The central value, $F_0 = 0.67$~GeV, is
only about 20\% higher than the value of $F$ obtained in the
nonrelativistic model.

Calculating the subleading terms $\lamtilde$, $G_1$ and $G_2$ is quite
similar to the nonrelativistic calculation.  The $\lamtilde$ term has to be
included by hand, just as before.  The hyperfine effect, $G_2$, which was
problematic in the nonrelativistic model, is also problematic in the
semirelativistic model.  Even though the one-loop potential, with its
milder $r\to0$ singularity, helped deal with the divergence at the origin
of $\phisr_\infty$, it does not alleviate the additional singularities in
the hyperfine potential.  The one-loop hyperfine potential
\cite{oneloop1,oneloop2},
\begin{eqnarray}
	V_\sw{HF} &=& \frac{32\pi}{3}\alpha_s(\mu)\left\{\left[1+
	 \frac{\alpha_s}{\pi}
	\left(\frac{5}{12}\beta_0 - \frac{11}{3} + \frac{15}{24}\ln
	\frac{m_Q}{m_q}\right)\right]\delta^3(\bold{r}) +
	\right.\\ \nonumber
	&&\left. \frac{\alpha_s}{\pi}
	\left(\frac{21}{8}-\frac{\beta_0}{4}\right)\left[\frac{1}{2\pi}
	\frac{1}{r^3} + 2\gamma_E\delta^3(\bold{r})\right]\right\},
	\label{eq:Vss}
\end{eqnarray}
contains terms just as singular as the tree-level hyperfine potential.
This means that even after the original divergence at the origin is
regulated, the hyperfine potential will introduce a new divergence.  This
result can be anticipated from the perturbative calculation in the
effective theory, where $G_2(\mu)$ gets a perturbative coefficient in
Eq.~(\ref{eq:fHQsubll}) beyond that of $F(\mu)$.  As a rule, terms which
have renormalization coefficients with non-trivial $\ln(\mu)$-dependence
diverge in the semirelativistic model.  $G_1$, which is protected from
renormalization by reparameterization invariance, does not diverge in the
model calculation.

Although the semirelativistic model $G_2$ calculation seems to be more
tractable than the similar problem in the nonrelativistic model, the
calculation is extremely sensitive to the small-$r$ dependence of the wave
function.  This is precisely where the numerical method breaks down, so the
calculation is not technically feasible.  Neither the nonrelativistic nor
semirelativistic models in present form give definite predictions for
$G_2$.  Fortunately, as argued in the previous section, indications are
that $G_2$ is negligibly small compared to $G_1$.

\begin{figure}
	\hfill\epsfxsize=0.5\hsize\epsffile{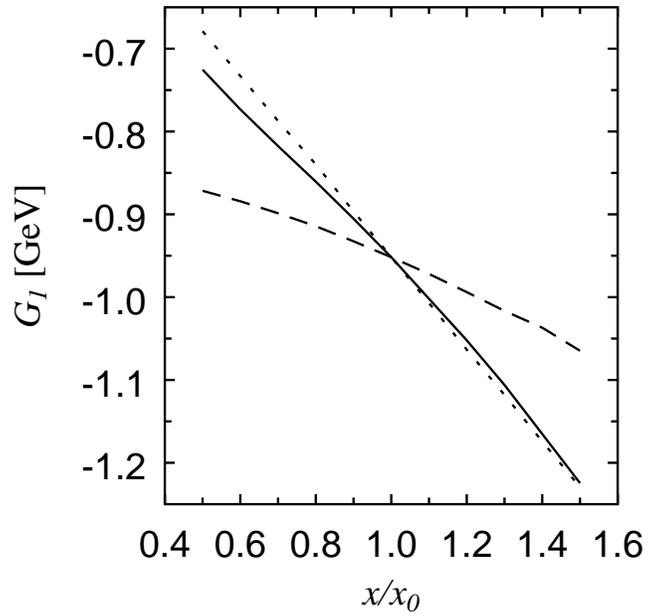}\hfill\hfill

	\caption{Parameter dependence of $G_1$ calculated using the
	nonrelativistic model.  The parameters ($x$) are $\Lambda$ (solid
	line), $m_q$ (dashed line) and $a$ (dotted line).}
\protect\label{fig:G1r}
\end{figure}

The kinetic energy term, $G_1$, can easily be calculated by treating the
$p^2/2m_Q$ term as a perturbation.  (The interpretation as a reduced mass
effect mentioned for the nonrelativistic model does not translate to the
semirelativistic model, so Eq.~(\ref{eq:nrG1rel}) no longer holds.) The
result is displayed in Fig.~\ref{fig:G1r}, once again showing dependence on
the various parameters.

Here the two models give dramatically different results.  $G_1$ is several
times larger in the semirelativistic model than it is the nonrelativistic
model.  Also the parameters are correlated such that the two models cannot
be made qualitatively similar by changing any combination of the
parameters.

\section{Discussion}
It is convenient to define a quantity $g_M$, such that
\begin{equation}
	f_M = F\left(1+\frac{g_M}{m_Q}\right) + {\cal O}
		\left(\frac{1}{m_Q^2}\right).
	\label{eq:defgM}
\end{equation}
\begin{table}
\caption{Comparison of the nonrelativistic and semirelativistic models.}
\protect\label{tab:the}
\begin{center}
\begin{tabular}{cccc}
\hline\hline
&$F$ ($F_0$) [GeV$^{3/2}$]&$G_1$ [GeV]&$g_P$ [GeV] \\
nonrelativistic & $0.55$ & $-0.14$ &$-0.31$\\
semirelativistic & $0.67$ & $-0.95$ &$-1.12$\\
\hline\hline
\end{tabular}
\end{center}
\end{table}
In Table~\ref{tab:the} I have summarized the results of the two model
calculations, including $g_M$ for pseudoscalar mesons, $g_P$.  (In general,
the pseudoscalar meson's $g_P$ will be different from the vector meson's
$g_V$ due to the effects of the hyperfine operator, $G_2$.  Since the
model calculations neglect the hyperfine contribution, $g_M = g_P = g_V$
for these calculations.) I have used $\lambar=m_q=0.33$~GeV to calculate
$g_P$. Note that $g_P$ in the nonrelativistic calculation is equal to the
naive guess from the introduction. This is actually fortuitous, because
the contribution of the $\lamtilde$ term, which accounts for half of the
value, is not included in the naive model. As stated in the introduction,
lattice calculations indicate that $g_P \approx 1$~GeV. The
semirelativistic calculation is consistent with that result, but the
nonrelativistic calculation is not. The difference is due to the $G_1$
contribution.

The nonrelativistic and semirelativistic values of $G_1$ differ not just
quantitatively, but qualitatively.  This qualitative difference would be
completely obscured by a model comparison done in the traditional way, {\em
i.e.,} by calculating only the decay constant and including the heavy quark
mass effects to all orders.  The heavy quark mass suppresses the effects
of the $G_1$ term in the decay constant itself.  The (heavy quark suppressed)
large difference at subleading order also tends to compensate the smaller
difference between the two models at leading order.

The origins of the discrepancy between the two models can be understood as
follows: For small $p$, the two Hamiltonians are the same.  For large $p$,
however, the kinetic energy term grows like $p^2$ nonrelativistically, but
only like $p$ relativistically.  This means that the semirelativistic
Hamiltonian is less confined in momentum space than the nonrelativistic
Hamiltonian, {\em i.e.,} the semirelativistic wave function is more spread
out in momentum space than the nonrelativistic wave function.  Because the
wave functions are normalized, an increase of the wave function at large
momentum must be compensated by a decrease at small momentum, so the
difference between the wave functions tends to cancel for $\psi(\vecr=0)$,
which can be written in $p$-space as
\begin{equation}
	\psi(\vecr = 0) = \int d^3\vecp\, \psi(\vecp).
	\label{eq:psi0p}
\end{equation}
$G_1$, however, is proportional to $\phi_\sw{KE}(\vecr=0)$, which can be
written as
\begin{equation}
	\phinr_\sw{KE}(\vecr) = \sum_{n\neq\infty} {\phinr_n(\vecr)\over E_n -
		E_\infty}\int
		d^3\vecp' \phi_n^*(\vecp') \frac{p^2}{2} \phinr_\infty(\vecp').
	\label{eq:kepertp}
\end{equation}
The $p^2$ factor in the integral emphasizes the large-$p$ differences in
the wave functions, making $G_1$ a sensitive probe of the large-momentum
tail of heavy-light wave functions.  This, in turn, leads to the following
rule: Quantities which are sensitive to the large-momentum shape of wave
functions are dramatically underestimated by the nonrelativistic quark
model.

This rule has implications for other processes.  In particular, it
indicates that nonrelativistic quark model calculations of processes at
large momentum transfer seriously underestimate the overlap of meson
wave functions.  An important example that has received much interest
lately is the process $B\to K^* \gamma$.  In the $B$ meson's rest frame the
$K^*$ has $\approx 1.3$~GeV of momentum, which is large compared to the
typical widths of meson wave functions in the nonrelativistic quark model.
This means that the overlap is dominated by the tails of the wave
functions, which I have just shown to be poorly described by the
nonrelativistic quark model.

This work not only provides an explanation for the conflict between the
nonrelativistic quark model and other estimates for the heavy-quark
symmetry breaking behavior in decay constants, it also suggests a
qualitative solution to an earlier conflict: In Ref.~\cite{me1}, I
calculated heavy-quark symmetry-violating corrections to form factors in
$B\to D^{(*)}l\nu$ transitions.  The predictions for the effects of the
heavy-quark kinetic energy operator in were an order of magnitude smaller
than a QCD sum rule calculation \cite{NeubertSLSR} of the same effect.
This work shows that the nonrelativistic quark model dramatically
underestimates the effect of the kinetic energy operator, in this case by a
factor of 6.  In the meantime, Neubert \cite{NeubertVir} has derived a
theorem showing the sum rule used in Ref.~\cite{NeubertSLSR} overestimate
the effects of the same operator.  While an explicit calculation is needed
for both, it appears that the two different types of models should now be
in qualitative agreement.

\section{Conclusions}

The nonrelativistic quark model provides a very simple picture of hadronic
physics.  While the picture is clearly too simple, it does yield insight
into the structure of hadrons.  Calculating decay constants in the
nonrelativistic quark model and a simple semirelativistic generalization
shows how the nonrelativistic quark model can work reasonably well overall,
yet fail to describe important details.  This calculation shows that the
nonrelativistic quark model does conflict with lattice and QCD sum rule
predictions for the size of heavy quark symmetry-breaking effects in
heavy-light decay constants.  However, this conflict can be removed by
going to a similar model with relativistic light quark dynamics.  Comparing
the two models shows that the nonrelativistic quark model should be
expected to fail for calculations which are sensitive to the large-momentum
tails of wave functions.

\section*{Acknowledgments}
I would like to thank M.G.\ Olsson for helpful discussions.  This research
was supported in part by the University of Wisconsin Research Committee
with funds granted by the Wisconsin Alumni Research Foundation, and in part
by the U.S.\ Department of Energy under grant DE-FG02-95ER40896.

\newpage


\newpage

\section*{Appendix}
This appendix describes the method I used to obtain the numerical results
in the text.  While the Schr\"odinger equation can easily be solved with a
wide variety of numerical techniques, the spinless Salpeter equation is
much more difficult.  The position-space representation of the spinless
Salpeter equation contains the problematic $\sqrt{-\nabla^2 + m_q^2}$
operator, while the momentum-space representation contains a complicated
convolution integral from the potential.

These problems can be avoided by using the Rayleigh-Ritz-Galerkin (RRG)
method \cite{Stakgold}, which easily handles both the Schr\"odinger and the
spinless Salpeter equations.  RRG is an extension of the elementary
variational method.  In the variational method, one chooses a state
parameterized by $\lambda$, then minimizes
\begin{equation}
	E_\sw{var} = \bra{\lambda}{\cal H}\ket{\lambda}
\end{equation}
with respect to $\lambda$.  For reasonable choices of $\ket{\lambda}$,
$E_\sw{var}^\sw{min}$ and $|{\lambda^\sw{min}}\rangle$ form good
approximations to the eigenvalue and eigenket, respectively.  In the RRG
method, one chooses an orthogonal set of $n$ vectors, $\ket{\lambda,i}$
$i=1\ldots n$, then minimizes
\begin{equation}
	E_\sw{RRG} = \bra{\Psi}{\cal H}\ket{\Psi},
	\label{eq:RRG}
\end{equation}
where $\ket{\Psi} = c_i\ket{\lambda,i}$.  One can calculate wave functions
and energy eigenvalues arbitrarily well by choosing sufficiently large $n$.
The problem is reduced to the numerically straightforward problems of
calculating integrals and solving a matrix equation for the $c_i$'s.

The difficulties with the spinless Salpeter equation can be avoided in the
RRG method by breaking up the expectation value of the Hamiltonian into
kinetic and potential pieces,
\begin{equation}
	{\cal H} = T + V,
\end{equation}
then writing Eq.~(\ref{eq:RRG}) as follows
\begin{equation}
	E_\sw{RRG} = \bra{\Psi}p\rangle\bra{p}T\ket{p}\langle p\ket{\Psi} +
		\bra{\Psi}x\rangle\bra{x}V\ket{x}\langle x\ket{\Psi} .
	\label{eq:RRG2}
\end{equation}
The resulting integrals are straightforward as long as the representations
of the $\ket{\lambda,i}$'s are known in both position and momentum space.

For the calculations in this work I used two different bases: the harmonic
oscillator basis and the confined pseudohydrogenic basis.  The former are
standard; the latter were developed in Ref.~\cite{Weniger} and first used
for the spinless Salpeter equation in Ref.~\cite{JaOlSu}.  All convergent
results are independent of basis.  The two different bases act as a
cross-check.  Since the spinless Salpeter wave function diverges at the
origin, the two methods do not agree in a small region around the origin.
The inset in Fig.~\ref{fig:semirel} shows the failure to converge in the
pseudohydrogenic basis.  The same plot with the harmonic oscillator basis
is different in the vicinity of the origin.  Nevertheless, the limiting
procedure in Eq.~(\ref{eq:F0def}) provides a finite quantity which is
basis-independent.  All the results in the text are independent of basis.

\end{document}